\documentclass[reqno]{amsart} \usepackage{amscd}
\usepackage{epsf}
\theoremstyle{remark} 
\numberwithin{equation}{section}
\newcommand{\field}[1]{\ensuremath{\mathbb{#1}}}
\newcommand{\CC}{\field{C}}

\begin{document}
\title[Quillen bundle and geometric prequantization]{Quillen bundle and Geometric Prequantization of Non-Abelian 
Vortices  on a Riemann surface}
\author{Rukmini Dey* and Samir K. Paul**}

\maketitle

\begin{abstract}
In this paper we prequantize the moduli space of non-abelian vortices. We 
explicitly calculate the symplectic form arising from the $L^2$ metric 
 and we construct a prequantum line bundle whose curvature is proportional to 
this symplectic form. The prequantum line bundle turns out to be Quillen's 
determinant line bundle with a modified Quillen metric. Next, as in the case 
of abelian vortices, we construct Quillen line bundles over the moduli space 
whose curvatures form a family of symplectic forms which are parametrised by 
$\Psi_0$, a section of a certain  bundle.
\end{abstract}         

\section{Introduction}

Geometric 
prequantization  of a symplectic manifold $({\mathcal M}, \Omega)$  is a construction of a prequantum line bundle 
${\mathcal P}$ 
 on ${\mathcal M}$ so that its curvature is proportional to the symplectic form.
The line bundle has to come equipped with a metric so that the Hilbert space of the prequantization is the space of  the square integrable sections of 
${\mathcal P}$.  
To every $f \in C^{\infty}({\mathcal M})$ we associate an operator acting on the Hilbert space, namely,  $\hat{f} = -i \hbar [\nabla_{X_f}^{\theta}] + f$ 
where $X_f$  is the vector field 
defined by $\Omega(X_f, \cdot) = - df$ and  $\theta$ is a $1$-form such that 
locally, $d \theta = \Omega$  and $\nabla_{X_f}^{\theta}$ is the covariant derivative with respect to $-\frac{i}{\hbar} \theta$ in the direction $X_f$. 
Then if $f_1, f_2 \in C^{\infty}({\mathcal M})$ and 
$f_3 = \{ f_1, f_2 \}_{\Omega}$, the Poisson bracket of the two induced by the symplectic form,  then $[\hat{f}_1, \hat{f}_2] = -i \hbar \hat{f}_3$, ~\cite{Wo}. 

The step from prequantization to quantization involves choice of a polarization and construction of operators which take polarised sections to polarised sections and  which satisfy the axioms of a deformation quantization, i.e. one may have to relax the condition that 
 $[\hat{f}_1, \hat{f}_2] = -i \hbar \hat{f}_3.$ Instead one might have 
  $[\hat{f}_1, \hat{f}_2] = -i \hbar \hat{f}_3 + o(\hbar^2) $. Construction of Toepliz operators out of projections to holomorphic sections of the prequatum line bundle (when the latter makes sense) and Berezin-Toeplitz deformation quantization has been carried out by Andersen in
  ~\cite{A1}, ~\cite{A2}.

The non-abelian vortices were first introduced in the literature perhaps by Bradlow ~\cite{Br}. 
The non-abelian vortex equations that we are considering were first studied 
in ~\cite{BDW}, ~\cite{BDGW} and subsequently studied by Baptista, ~\cite{B}. Let $M$ be a compact Riemann 
surface and let $\omega = h^2 dz \wedge d \bar{z}$ be the purely imaginary 
volume form on it, (i.e. $h$ is real). 
Let $V$ be a complex vector bundle associated to $P.$ Let  $A$ be a unitary connection on a $V$
 i.e. $A$ is a $1$-form such that $A^* = -A,$ and $A= A^{(0,1)} + A^{(1,0)}$ 
i.e. $   A^{(0,1)*} = -A^{(1,0)}$. 
Let $E = \oplus^N V $ 
 i.e. direct sums of $N$ copies of $V$ 
(with $N$ being the rank of $V$).  Let $\Psi$ be a section of $E$, i.e. 
$\Psi \in  \Gamma(M,E)$. In this case the section $\Psi$ can 
be regarded locally as a function on $M$ having values on the $N\times N$ 
matrices. The hermitian conjugate $\Psi^*$ is defined w.r.t. 
 a Hermitian metric $H$ on $E$, so that in a  unitary trivialization of $E$ it 
is represented by the Hermitian conjugate matrix of $\Psi$. 

The pair $(A, \Psi)$ will be said to satisfy the non-abelian vortex equations if

\hspace{1in} $(1)$ $ \rm{\;\;\;\;\;}$ $ F(A) = (\tau I -\Psi \Psi^*) \omega,$

\hspace{1in} $(2)$ $\rm{\;\;\;\;\;}$ $\bar{\partial}_A \Psi = 0,$

where $F(A)$ is the curvature of the connection $A$ and $d_A = \partial_A + \bar{\partial}_A $ is the decomposition of the covariant derivative operator
into $(1,0)$ and $(0,1)$ pieces.  Note we take $e=1$ in ~\cite{B}. 
Let ${\mathcal S}$ be the space of solutions to $(1)$ and $(2)$.
 There is a gauge group $G$ acting on the space of $(A, \Psi)$ which leaves the equations invariant. We take the group $G$ to be $U(n)$  and locally it looks like ${\rm Maps} (M, U(n)).$ If $g$ is an $U(n)$ gauge transformation then 
$(A_1, \Psi_1)$ and $(A_2, \Psi_2)$ are gauge equivalent if 
$A_2 = g^{-1}dg + g^{-1}A_1g $ and $\Psi_2 = g^{-1} \Psi_1$. 
  Taking the quotient by the gauge group of ${\mathcal S}$ gives  the moduli 
space of solutions to these 
equations and is denoted by ${\mathcal M}$. In the case where $E$ is of this 
special form, ${\mathcal M}$ is known to be a smooth manifold for big values of $\tau$ , more precisely when $\tau > 2 \pi d / n Vol (\Sigma)$ (see references in ~\cite{B}) . 
It is known that   there is a natural metric on the moduli space 
${\mathcal M}$ and in fact the metric is K\"{a}hler.
In this paper, we show the metric explicitly and write down the symplectic 
(in fact, the K\"{a}hler form ) $\Omega$ arising from this metric and the 
complex structure.

As done in the case of abelian vortex moduli space by  Dey, ~\cite{D2},   we show here that there exists a holomorphic 
prequantum line bundle, namely, a determinant line bundle, which has a modified Quillen metric such that the  Quillen 
curvature is proportional to the sympletic form $\Omega.$ This method was first used in constructing determinant line bundles over stable triples by Biswas and Raghavendra, ~\cite{BR}.
Also, we should mention that 
Biswas and Schumacher, ~\cite{BS},  has used the Quillen metric on moduli spaces of coupled vortex equations on  a complex projective variety and shown that the K\"{a}hler form is the Chern form of a Quillen metric on a certain determinant line bundle.

 Next, in this paper, as in ~\cite{D1}, we construct line bundles over the 
moduli space  whose curvatures are proportional to  a family of symplectic 
forms parametrised by $\Psi_0$, a section of  $E \otimes K$.

In the general case, when $E$ is an arbitrary vector bundle associated to a 
$U(n)$ principal bundle $P$
the moduli space of non-abelian vortices is not known to be non-empty or smooth.However, 
the calculations done in this paper will still be valid outside the singular 
locus.

Note: In this paper, many details of proofs  are skipped because they coincide 
with  ~\cite{D1}, ~\cite{D2} and an expository paper, ~\cite{DP}.

\section{Metric and symplectc forms}

Let ${\mathcal A} $ be the space of all unitary connections on $P$ and 
$\Gamma (M, E)$ be sections of $E$. 
Let ${\mathcal C} = {\mathcal A} \times \Gamma (M, E)$ be the configuration 
space on which 
equations $(1)$ and $(2)$ are imposed. Let $p= (A, \Psi) \in {\mathcal C}$, $X
= ( \alpha_1, \beta)$, $Y= (\alpha_2, \eta)$
$\in T_p {\mathcal C} \equiv \Omega^1(M, u(n)) \times \Gamma(M, E) $ i.e.
 $\alpha_i = \alpha_i^{(0,1)} + \alpha_i^{(1,0)}$ such that 
$\alpha_i^{(0,1)*}= - \alpha_i^{(1,0)}, i = 1,2.$
On ${\mathcal C}$ one can define a metric
\begin{eqnarray*}
 {\mathcal G} ( X, Y) = \int_M {\rm Tr} (*_1\alpha_1 \wedge \alpha_2)  + 2i \int_M  {\rm Tr} (\frac{\beta \eta^* + \beta^* \eta}{2}) \omega
\end{eqnarray*}
and an almost complex structure  ${\mathcal I} = \left[
\begin{array}{cc}
*_1 & 0  \\
0 & i 
\end{array} \right] : T_p {\mathcal C} \rightarrow T_p {\mathcal C}$
where   $*_1: \Omega^{1} \rightarrow \Omega^{1}$ is  the Hodge star
operator on $M$ such that $*_1 (\alpha^{(1,0)}) = - i \alpha^{(1,0)}$ and $*_1 \alpha^{(0,1)}= i \alpha^{(0,1)}$ (i.e. $(0,1)$ forms are holomorphic 
w.r.t. this.) 

It is easy to check that  ${\mathcal G}$ is the $L^2$ metric on ${\mathcal C}$.

\subsection{The symplectic forms  $\Omega$ and $\Omega_{\Psi_0}$}

We define
\begin{eqnarray*}
\Omega(X, Y) &=& -\int_{M} {\rm Tr} (\alpha_1 \wedge \alpha_2) + 2i \int_{M}  {\rm Tr} (\frac{i \beta  \eta^* - i \beta^* \eta}{2}) \omega\\ 
&=& -\int_{M} {\rm Tr} (\alpha_1 \wedge \alpha_2) - \int_{M}  {\rm Tr}( \beta  \eta^* - \beta^*  \eta ) \omega
\end{eqnarray*}
such that $ {\mathcal G} ({\mathcal I} X, Y) = \Omega ( X, Y).$

Let $\zeta \in {\rm Maps} (M, u(n))  $ be the Lie algebra of the
gauge group (the gauge group element being $g = e^{ \zeta}$ ); note that 
$\zeta^* = - \zeta$ .  
It generates a vector field $X_{\zeta}$ on ${\mathcal C}$ as follows :
$$X_{\zeta} (A, \Psi) = (d \zeta, -\zeta \Psi) \in T_p
{\mathcal C}$$ where $ p = (A, \Psi) \in {\mathcal C}.$

We show next that $X_{\zeta}$ is Hamiltonian. Namely, define
$H_{\zeta} : {\mathcal C} \rightarrow {\CC} $ as follows: $$
H_{\zeta} (p) = \int_{M} {\rm Tr} [\zeta \cdot (  F_{A} -( 1-\Psi \Psi^*) \omega)]. $$  
Then for $X = (\alpha, \beta) \in T_p {\mathcal C}$,
 \begin{eqnarray*}
 dH_{\zeta} ( X ) & = & \int_M {\rm Tr} (\zeta d \alpha)  + \int_M {\rm Tr} [\zeta   
( \Psi  \beta^* + \beta \Psi^*  )]  \omega    \\
 &= &- \int_M {\rm Tr} [(d \zeta) \wedge \alpha]  -  \int_M {\rm Tr}[\beta  \Psi^* (-\zeta) - \beta^*  \zeta \Psi]   \omega  \\
 &= &- \int_M {\rm Tr} [(d \zeta) \wedge \alpha]  -  \int_M {\rm Tr}[\beta  (\zeta \Psi)^*  - \beta^*  (\zeta \Psi)]   \omega  \\
&= &- \int_M {\rm Tr} [(d \zeta) \wedge \alpha]  -  \int_M {\rm Tr}[  (-\zeta \Psi) \beta^*-   (-\zeta \Psi)^* \beta]   \omega  \\
& = & \Omega ( X_{\zeta},  X ),
 \end{eqnarray*}
where we use that $\zeta^* = - \zeta$.

 Thus we can define the moment map $ \mu : {\mathcal C} \rightarrow
 \Omega^2 ( M, u(n) )= {\mathcal G}^* $ ( the dual of the Lie
 algebra of the gauge group)  to be $$ \mu ( A, \Psi)
 \stackrel{\cdot}{=} (F(A) - ( 1-\Psi  \Psi^*)  \omega). $$ Thus equation 
$(1)$ is $\mu = 0$.

It can be shown exactly along lines of Dey, ~\cite{D1} that $\Omega $, 
 ${\mathcal G}$ and ${\mathcal I}$ descend to ${\mathcal M}$ so that  the latter is symplectic and almost complex.

Next we will define a family of symplectic forms on ${\mathcal M}.$

Let $\Psi_0 \in \Gamma(M, E \otimes K)$ where $K$ is the canonical bundle such that $\Psi_0$ has zero only in a set of measure zero. 

\begin{eqnarray*}
 \Omega_{\Psi_0} (X, Y) 
&=&  - [\int_M {\rm Tr} (\alpha_1 \wedge \alpha_2) - \int_M {\rm Tr} [(\eta^* \beta - \beta^* \eta) \Psi^*_0 \wedge \Psi_0] \\
&=&     - [\int_M {\rm Tr} (\alpha_1 \wedge \alpha_2) + \int_M {\rm Tr} \{(\eta^* \beta - \beta^* \eta) f(\Psi_0) \} \omega ]
\end{eqnarray*}
where $f(\Psi_0) \bar{\omega}  = \Psi^*_0 \wedge \Psi_0$ is zero only in a set of measure zero, $f(\Psi_0)$ being the matrix $A^* A$ where $\Psi_0 = A dz$. 
 We used the fact that ${\rm Tr} (\Psi_0 \beta^* \eta \wedge \Psi_0^*) = - {\rm Tr}( \beta^* \eta \Psi_0^* \wedge \Psi_0) $  has to cross over a $\Psi_0^*$ which is a matrix-valued $(0,1)$ form.

This is non-degenerate and is a symplectic form on ${\mathcal M}.$

This follows from the fact that 
\begin{eqnarray*}
\Omega_{\Psi_0} ( {\mathcal I} (\alpha_1, \beta) , (\alpha_1, \beta)) =  -4\int_M  |a|^2 dx \wedge dy  - 4\int_M {\rm Tr} (\beta^* \beta A^* A)    h^2 dx \wedge dy
 \end{eqnarray*}
where $\omega = -2i h^2 dx \wedge dy $ and 
$\alpha_1 = a dz - a^* d \bar{z}  \in \Omega^1(M, u(n))$ 
and $*_1\alpha_1 = -i( a dz + a^* d \bar{z} )$. 
 ${\rm Tr} (\beta^* \beta A^* A) $ is positive definite because it is 
${\rm Tr} (A \beta^* \beta A^* ) = {\rm Tr} ((A \beta^*) (A \beta^*)^*)$ which is of the form ${\rm Tr} (C C^*)$ which is obviously positive definite outside 
maybe a set of measure zero.

\section{Prequantum line bundle}
 In this section we briefly review the Quillen construction of the determinant 
line bundle of the Cauchy Riemann operator  $\bar{\partial}_A = \bar{\partial} + A^{(0,1)}$, ~\cite{Q},
which  enables  us to construct prequantum line bundle on the vortex moduli 
space.

\subsection{Determinant line bundle of Quillen}

First let us note that a connection $A$ on a $U(n)$-principal bundle induces 
a  connection on any associated line bundle $E$. 
We will denote this connection also by $A$ since  the same ``
Lie-algebra valued $1$-form'' $A$ (modulo representations)  gives  a covariant 
derivative operator enabling you to take derivatives of  sections of $E$.

A very clear description of
the determinant line bundle can be found in ~\cite{Q} and
~\cite{BF}. We also give more details in ~\cite{DP}.

${\mathcal A} =$ space of unitary connections on a vector bundle $E$ 
associated to
a principal $G$ bundle on a Riemann surface. 
$A = A^{1,0} + A^{0,1}$ with $A^{1,0*} = - A^{0,1}$. Thus identify 
${\mathcal A} = {\mathcal A}^{0,1}$. 

Construct a line bundle ${\mathcal L}$ on ${\mathcal A}^{0,1}$ as follows. 
The fiber on top of $A^{0,1}$ is $${\rm det} (\bar{\partial}_A) = \wedge^{\rm{top}} ({\rm Ker} \bar{\partial}_A)^{*} \otimes \wedge^{\rm{top}}({\rm Coker}
\bar{\partial}_A)$$ 

Quillen's ingenious construction: ${\mathcal L}$ carries a metric and a
connection s.t. the curvature is exactly $\Omega(\alpha, \beta) =
\int_{\Sigma} {\rm Tr} (\alpha \wedge \beta) $ on ${\mathcal A}^{0,1}$.

Let $$\Delta_A = *\bar{\partial}_A * \bar{\partial}_A$$ be the Laplacian of
the $\bar{\partial}_A$ operator.

$K^a_0 = $ sum of eigenspaces of $\Delta_A$ for eigenvalues less than $a$ and 
$\bar{\partial}_A K^a_0 = K^a_1$.

Consider the exact sequence

$0 \rightarrow {\rm Ker} \bar{\partial}_A \rightarrow K^a_{0} \stackrel{\bar{\partial}_A}{\rightarrow} K^a_{1}
\rightarrow {\rm Coker} \bar{\partial}_A \rightarrow 0$

$\lambda = \wedge^{top} ({\rm Ker} \bar{\partial}_A)^* \otimes \wedge^{top}
({\rm Coker}
\bar{\partial}_A)$ can be identfied with 
$\lambda^a = \wedge^{top} (K_0^a)^* \otimes \wedge^{top} (K_{1}^a)$
 over the open set
$U^a = \{ a \notin {\rm spec} \Delta_A \} \subset {\mathcal A}$.

$\lambda^a$ is a smooth line bundle over $U^a$. For if $a, b \notin {\rm spec}
\Delta_A$, $a<b$, $K_0^{(a,b)}=$ union of eigenspaces of $\Delta_A$
corresponding to eigenvalues $\mu$ within $a < \mu <b$.

Let $ K_1^{(a,b)} = \bar{\partial}_A ( K_0^{(a,b)})$. 

$\lambda^{(a,b)} = \wedge^{top} (K_0^{(a,b)})^* \otimes \wedge^{top}
(K_{1}^{(a, b)})$ over $U^a \cap U^b$ 

Let $\bar{\partial}_A^{(a,b)} = \bar{\partial}_A|_{K_0^{(a,b)}}$ then
$\lambda^b = \lambda^a \otimes \lambda^{(a,b)}$.

The identification of $\lambda^a$ and $\lambda^b$ via $\lambda$ is given by
the mapping $s \in \lambda^a \rightarrow s \otimes
{\rm det}(\bar{\partial}_A^{(a,b)}) \in \lambda^b$.

Now under the gauge transformation,
$\bar{\partial}_A = \bar{\partial} + A^{(0,1)} \rightarrow g (\partial +
A^{0,1}) g^{-1}$,  $\Delta_g = g \Delta_A g^{-1}$. 

There is an isomorphism
of eigenspaces $s \rightarrow gs $ (with the same eigenvalues). Thus when one 
identifies

$\lambda = \wedge^{top} ({\rm Ker} \bar{\partial}_A)^* \otimes \wedge^{top}
({\rm Coker}
\bar{\partial}_A)$  with 

$ \lambda^a = \wedge^{top} (K_0^a)^* \otimes \wedge^{top} (K_{1}^a)$
there is an isomorphism of fibers over $U^a$:

$\wedge^{top} (K_0^a(\Delta_A))^* \otimes \wedge^{top} (K_{1}^a(\Delta_A))
\equiv \wedge^{top} (K_0^a(\Delta_{A_g}))^* \otimes \wedge^{top}
(K_{1}^a(\Delta_{A_g}))$.
The fiber over $U^a/ G$ is the equivalence class of this fiber.

Like this we can define the line bundle on ${\mathcal A}/ G.$

\subsection{Quillen metric}

Using the Hermitian structure on $E$ (the vector bundle on the Riemann
surface $\Sigma$) and therefore the one on $\Omega^{p,q}(E)$ one
can define Hermitian metrics on $K^a_{0}$ ($K^a_{1}$) over $U^a$ and 
$K_{0}^{(a,b)} $ ($K_{1}^{(a,b)} $)over $U^a \cap U^b.$ The bundles $\lambda^a$ and $\lambda^{(a,b)}$ are then
naturally endowed with metric $| \cdot |^a$ and $|\cdot |^{(a,b)}$. 

Over $U^a \cap U^b,$ $|s \otimes {\rm det} \bar{\partial}_A^{(a,b)}|^b =
|s|^a| {\rm det} \bar{\partial}_A^{(a,b)}|^{(a,b)}$

When identifying $\lambda$ with $\lambda^a$ or $\lambda^b,$ the metrics $|\cdot
|^a$ and $|\cdot |^b$ are related to each other by 

$|\cdot |^b = |\cdot |^a |{\rm det} \bar{\partial}_A^{(a,b)}|^{(a,b)}$.

To correct this discrepancy, Quillen does a zeta function regularisation:

Let $\zeta_A^a(s)$ is exactly the zeta function of the operator $\Delta_A$
restricted to eigenspaces whose eigenvalues are larger than $a$.

Then for $0 < a < b < \infty$ we can also define $\zeta_A^{(a,b)}(s).$ Clearly,
$\zeta_A^a(s) = \zeta_A^{(a,b)}(s) + \zeta_A^b(s).$

Also, 
$ | {\rm det} \bar{\partial}_A^{(a,b)}|^{(a,b)} = {\rm exp} \{ -\frac{1}{2} \frac{\partial
\zeta_A^{(a,b)}}{\partial s} (0) \} = {\rm exp} \{ -\frac{1}{2} 
\zeta_A^{\prime (a, b)} (0) \}. $

Thus one can define $||\cdot||^a $ to be the metric on $\lambda^a$ which is such
that if $l \in \lambda^a$ 

$ || l ||^a = |l|^a  {\rm exp} \{ -\frac{1}{2} \frac{\partial
\zeta_A^a}{\partial s} (0) \} = |l|^a  {\rm exp} \{ -\frac{1}{2} 
\zeta_A^{\prime a} (0) \}. $

Thus under the canonical identification of $\lambda$ with $\lambda^a$ over
$U^a,$ the metrics $||\cdot||^a$ patch into a smooth metric $||\cdot||$ on $\lambda.$

\subsection{The curvature formula}

This metric induced a connection and Quillen computed the curvature of the determinant
line bundle $\lambda.$ For details, see ~\cite{Q}, ~\cite{DP}.
In fact the K\"{a}hler potential for this symplectic form is proportional to 
 $\zeta_A^{\prime } (0)  $ 

The curvature turned out to be  proportional to
$$\frac{i}{2 \pi}\Omega_0(\alpha,
\beta) = - \frac{i}{2 \pi} \int_{\Sigma} {\rm Tr} (\alpha \wedge \beta),$$ the symplectic form, ~\cite{Q}.

Thus we are in the situation for geometric prequantization of the affine space 
${\mathcal A}$.

\section{Prequantum bundle on ${\mathcal M}$}

We first define the Quillen's determinant bundle ${\mathcal P} = {\rm det}(\bar{\partial}_A)$ on the affine space 
${\mathcal C} = {\mathcal A} \times \Gamma(M, E),$ i.e. over each point $(A, \Psi)$ we define the fiber as that of 
${\rm det}(\bar{\partial}_A)$ independent of $\Psi$. However we modify the Quillen metric to $e^{- \zeta_A^{\prime}(0) - \frac{i}{2 \pi} \int_M {\rm Tr} ( \Psi \Psi^*) \omega}$, which now depends on both $A$ and $\Psi$.  

This descends to ${\mathcal C}/ G$ 
because  under the gauge transformation,
$\bar{\partial}_{A} = \bar{\partial} + A^{(0,1) } \rightarrow g (\partial +
A^{0,1} ) g^{-1}$,  $\Delta_g = g \Delta_{A} g^{-1}.$

There is an isomorphism
of eigenspaces $s \rightarrow gs $ (with the same eigenvalues). Thus when one 
identifies

$\lambda = \wedge^{top} ({\rm Ker} \bar{\partial}_{A})^* \otimes \wedge^{top}
({\rm Coker}
\bar{\partial}_{A})$  with 

$ \lambda^a = \wedge^{top} (K_0^a)^* \otimes \wedge^{top} (K_{1}^a)$
there is an isomorphism of fibers over $\tilde{U}^a = U^a \times V, $
where $U^a$ is as before an open set containing $A$ and $V$ is an open set in $\Gamma(M, E)$  containing $\Psi$.

$\wedge^{top} (K_0^a(\Delta_{A}))^* \otimes \wedge^{top} (K_{1}^a(\Delta_{A}))
\equiv \wedge^{top} (K_0^a(\Delta_{(A,g)}))^* \otimes \wedge^{top}
(K_{1}^a(\Delta_{(A, g)}))$.
The fiber over $\tilde{U}^a/G$ is the equivalence class of this fiber.

Like this we can define the line bundle on ${\mathcal C} 
/ G$.

Then we restrict it to the moduli space ${\mathcal M} \subset 
{\mathcal C} 
/G$.

{\bf Curvature and symplectic form:}

 Following ~\cite{BR}, we  give ${\mathcal P}$  a modified Quillen metric, namely, we multiply the Quillen metric $e^{-\zeta_A^{\prime}(0)}$ by the factor  $e^{- \frac{i}{2 \pi} \int_{M} {\rm Tr}  (\Psi \Psi^*) \omega} ,$ where recall $\zeta_A(s)$ is the $zeta$-function 
corresponding to the Laplacian of the $\bar{\partial} + A^{0,1}$ operator. We calculate the curvature for this modified metric on the affine space. 
The $zeta$ part of the metric contributes  $- \frac{i}{2 \pi} \int_{M} {\rm Tr} (\alpha_1 \wedge \alpha_2)$ to the curvature $\Omega$ as is well known, ~\cite{Q}.
 The second part   $e^{- \frac{i}{2 \pi} \int_{M}  {\rm Tr} (\Psi \Psi^*) \omega} ,$ contributes to the second part of the curvature form $\Omega$, namely, 
$- \frac{i}{2 \pi} \int_{M} {\rm Tr} (\beta  \eta^* - \eta^* \beta) \omega$ as follows:

Let $N = \int_{M} {\rm Tr} (\Psi(z, \bar{z})  \Psi^* (z, \bar{z})) \omega(z, \bar{z})$, where 
$\omega(z, \bar{z})$ is the volume form on the Riemann surface.

\begin{eqnarray*}
\tau &=& \int_M [\int_M {\rm Tr} [\frac{\delta^2 (\Psi(z, \bar{z}) \Psi^*(z, \bar{z})) }{\delta \Psi(z^{\prime}, \bar{z}^{\prime} ) \delta \Psi^* (z^{\prime}, \bar{z}^{\prime})} \delta \Psi( z^{\prime}, \bar{z}^{\prime} ) \wedge \delta \Psi^*(z^{\prime}, \bar{z}^{\prime})] \omega(z^{\prime}, \bar{z}^{\prime}) ] \omega(z, \bar{z}) \\
&=& \int_M  [\int_{M}{\rm Tr} [ \delta(z - z^{\prime})  \delta(\bar{z} - \bar{z^{\prime}})  (\delta \Psi(z^{\prime}, \bar{z}^{\prime}) , \wedge \delta \Psi^* (z^{\prime} , \bar{z}^{\prime})) \omega(z^{\prime},\bar{z}^{\prime})]] \omega(z, \bar{z})  \\
&=& \int_{M} {\rm Tr} [\delta \Psi(z, \bar{z}) \wedge \delta \Psi^*(z, \bar{z})] \omega(z, \bar{z})
\end{eqnarray*}

Then $\tau$ is a two form on the affine space $\Gamma(M, E)$ such that 
$\tau (\beta, \eta) = \int_{M} {\rm Tr} (\beta \eta^* - \eta^*  \beta) \omega$.

The addition to the curvature two form due to the modification of the metric by $exp(- \frac{i}{ 2\pi} N)$ is $\partial_{\Psi} \partial_{\Psi^*}  log(exp(-\frac{i}{2 \pi} N)) =\partial_{\Psi} \partial_{\Psi^*}  (-\frac{i}{2 \pi} N )  $ which is $- \frac{i}{2 \pi} \tau$.

Thus the curvature of ${\mathcal P}$ with the modified Quillen metric is indeed \begin{eqnarray*}
\frac{i}{ 2 \pi} \Omega((\alpha_1, \beta), (\alpha_2, \eta)) =  - \frac{i}{2 \pi} [\int_{M} {\rm Tr} (\alpha_1 \wedge \alpha_2) + \int_{M}{\rm Tr} ( \beta \eta^* - \beta^* \eta )  \omega]
\end{eqnarray*}
the symplectic form on ${\mathcal M}$

{\bf Polarization:} In passing from prequantization to quantization, one needs 
a polarization. It can be shown that the almost complex structure 
${\mathcal I}$ is integrable on ${\mathcal M},$ ~\cite{B}.
   In fact, $\Omega$ is a K\"{a}hler form and 
${\mathcal G} (X,Y) = \Omega(X, {\mathcal I}Y)$
is a K\"{a}hler metric on the moduli space (since it is positive definite).
 ${\mathcal P}$ is a holomorphic line bundle on ${\mathcal M}$. 
Thus we can take holomorphic square integrable sections of 
${\mathcal P}$ as our Hilbert space.
The dimension of the Hilbert space is not easy to compute. (For instance,
the holomorphic sections of the determinant line bundle on the moduli space of  flat connections for $SU(2)$ gauge group is the Verlinde dimension of the 
space of  conformal blocks in a 
certain conformal field theory). This would be a topic for future work.

\section{Alternate way of prequantizing the moduli space}

Here we modify the method used in ~\cite{D1}, ~\cite{D2} to accommodate non-abelian 
vortices.

\subsection{Modified determinant line bundles on the moduli space}

Let us modify the determinant line bundle construction on a different affine 
space.

Let ${\mathcal C_{+}} = {\mathcal A} + {\mathcal B}$ where ${\mathcal A}$ is the 
space of connections (its $(0,1)$ part only) on $E$, as in ~\cite{Q}, and 
${\mathcal B}$  subspace of $\Omega^1(\Sigma, u(n)) $ ,i.e. a subspace of the Lie-algebra valued $(0,1)$ forms which transforms as 
$B_g = gBg^{-1}$. $B = B^{(1,0)} + B^{(0,1)} $ where  $B^{(1,0)*} = - B^{(0,1)}$ 

We define  coordinates on this affine space as 
$w= A^{0,1} + B^{0,1}$  which is holomorphic with respect to the usual Hodge star operator which takes $*(\alpha^{(0,1)}+ \beta^{(0,1)}) = i (\alpha^{(0,1)} + \beta^{(0,1)})$ and  $*(\alpha^{(1,0)}+\beta^{(1,0)}) = -i (\alpha^{(1,0)} + \beta^{(1,0)}).$

Let the Cauchy-Riemann operator $\bar{\partial}_{(A,B, +)}$ be defined locally on the Riemann surface $\Sigma$ as $\bar{\partial} + A^{0,1} + B^{0,1}$  and it differentiates sections of $E$. Note if $B=0$, we get back Quillen's construction, 
~\cite{Q}.

One can show that one can   define the line bundle  $$ {\mathcal L}_{+} = {\rm det} \bar{\partial}_{(A,B, +)} = \wedge^{\rm{top}} ({\rm Ker} \bar{\partial}_{(A,B, +)})^{*} \otimes \wedge^{\rm{top}}({\rm Coker}
\bar{\partial}_{(A,B, +)})$$ on ${\mathcal C_{+}}$, and subsequently on $ ({\mathcal A} \times {\mathcal B})/ G $, see ~\cite{DP} for details.

The curvature corresponding to this line bundle is 

$$ -\frac{i}{2 \pi} \int_{\Sigma}  {\rm Tr} [(\alpha_1 + \beta_1) \wedge (\alpha_2 + \beta_2)].$$

Exactly analogous to the case of ${\mathcal C_+}$, we define the line bundle 
on ${\mathcal C_-} = {\mathcal A} - {\mathcal B}$ to be 
 $${\mathcal L}_{-} = {\rm det} \bar{\partial}_{(A,B, -)} = \wedge^{\rm{top}} ({\rm Ker} \bar{\partial}_{(A, B,-)})^{*} \otimes \wedge^{\rm{top}}({\rm Coker}
\bar{\partial}_{(A,B, -)})$$ where now  
$\bar{\partial}_{(A,B, -)} = \bar{\partial} + A^{0,1} - B^{0,1}$. As described in ~\cite{DP} it is easy to define it on 
 $({\mathcal A} \times {\mathcal B}) 
/ G.$

The curvature formula now becomes:

$$ -\frac{i}{2 \pi} \int_{\Sigma}  {\rm Tr} [(\alpha_1 - \beta_1) \wedge (\alpha_2 - \beta_2)]$$

\subsection{Special values of $B$}
Let $\Psi_0 \in \Gamma( M, E \otimes K)  $ where $K$ is the canonical bundle such that $\Psi_0$ has  zeroes only in  a set of measure zero on $M$.

 Now let us take 
$B= B^{(0,1)} + B^{(1,0)} =  \Psi \Psi^*_0 - \Psi_0 \Psi^*$

Then ${\mathcal L}_+ $ defined on $ ({\mathcal A} \times {\mathcal B}) 
/ G$ will have curvature:

$ \frac{-i}{2 \pi} \int_M {\rm Tr} [(\alpha_1 + \beta \Psi_0^* - \Psi_0 \beta^*)
\wedge (\alpha_2 + \eta \Psi_0^* - \Psi_0 \eta^*)]$

and ${\mathcal L}_- $  also defined on   $ ({\mathcal A} \times {\mathcal B}) 
/ G $  will have curvature:

$ \frac{-i}{2 \pi} \int_M {\rm Tr} [(\alpha_1 - \beta \Psi_0^* + \Psi_0 \beta^*)
\wedge (\alpha_2 - \eta \Psi_0^* + \Psi_0 \eta^*)]$

An easy calculation shows that ${\mathcal P}_{\Psi_0} =  {\mathcal L}_+  \otimes {\mathcal L}_- $ has curvature, by adding the above two,  $\frac{i}{ \pi} \Omega_{\Psi_0} (X, Y),$ see section $(2.1)$.

As described in ~\cite{DP}, the line bundle ${\mathcal P}_{\Psi_0}$ is well 
defined on $({\mathcal A} \times {\mathcal B})/G$ and hence on 
$({\mathcal A} \times \Gamma (M, E))/G$ and hence on ${\mathcal M} \subset ({\mathcal A} \times \Gamma (M, E))/G $ and the curvature formula descends to the moduli space.

As was mentioned before  $\Omega_{\Psi_0}$ is a family of symplectic forms 
parametrised by $\Psi_0$.

{\bf Remark:}
Thus we have two candidates for the Hilbert space of quantization, namely, 
holomorphic sections of ${\mathcal P}$ and ${\mathcal P}_{\Psi_0}$. It would be
interesting to see if they give equivalent quantization.

* School of Mathematics, Harish Chandra Research Institute, Jhusi,
Allahabad, 211019, India. email: rkmn@mri.ernet.in 
email: rukmini.dey@gmail.com

**S.N. Bose National Centre for Basic Sciences, Sector-III, Block - JD, Salt Lake, Kolkata - 700 098, India. email: smr@bose.res.in

\end{document}